\documentclass[submit]{smj}
\usepackage[utf8x]{inputenc}
\usepackage{amsmath, amssymb, amsthm, nccmath}
\usepackage{tablefootnote}
\usepackage{scrextend}
\usepackage{booktabs}
\usepackage[dvipsnames]{xcolor}

\Author{Antonello Maruotti\Affil{1,}\Affil{2}, 
        Luca Merlo\Affil{3} 
        and Lea Petrella\Affil{4}
}
\AuthorRunning{Antonello Maruotti \textrm{et al.}} 

\Affiliations{
\item Department of Mathematics, University of Bergen, Bergen, Norway
\item Department of Law, Economics, Political Sciences and Modern Languages, LUMSA University, Rome, Italy
\item Department of Statistical Sciences, Sapienza University of Rome, Rome, Italy
\item MEMOTEF Department, Sapienza University of Rome, Rome, Italy

}   

\CorrAddress{Luca Merlo, Department of Statistical Sciences, Sapienza University of Rome, Piazzale Aldo Moro 5, 00185 Rome, Italy}
\CorrEmail{luca.merlo@uniroma1.it}
\CorrPhone{(+39) 06 49911}
\CorrFax{(+39) 06 49911}

\Title{A two-part finite mixture quantile regression model for semi-continuous longitudinal data}
\TitleRunning{A two-part finite mixture quantile regression model}

\Abstract{
This paper develops a two-part finite mixture quantile regression model for semi-continuous longitudinal data. 
The proposed methodology allows heterogeneity sources that influence the model for the binary response variable, to influence also the distribution of the positive outcomes. As is common in the quantile regression literature, estimation and inference on the model parameters are based on the Asymmetric Laplace distribution. Maximum likelihood estimates are obtained through the EM algorithm without parametric assumptions on the random effects distribution. In addition, a penalized version of the EM algorithm is presented to tackle the problem of variable selection. The proposed statistical method is applied to the well-known RAND Health Insurance Experiment dataset which gives further insights on its empirical behavior.
}


\begin{document}

\maketitle

\section{Introduction}\label{sec:intro}
Semi-continuous data are non-negative and characterized by the coexistence of high kurtosis, positive skewness and an abundance of zeros observed often enough that there are compelling substantive and statistical reasons for special treatment (\cite{belotti2015twopm}). Modeling non-negative data with clumping at zero has a long history in the statistical literature and several competing models have been proposed (see \cite{min2002modeling} and \cite{min2005random} and references therein). Such data structure arises on many occasions: economics, actuarial sciences, environmental modeling and health services (see \cite{liu2009joint, belasco2012modelling, neelon2016modeling1, neelon2016modeling2} and \cite{farewell2017two}).
Linear regression approaches can be considered as starting points of the analysis (see e.g. \cite{Iversen2015}). Nevertheless, parameter estimates are sensitive to extreme values and likely to be inefficient if the underlying distribution is not Gaussian (\cite{Zhou2002} and \cite{Basu2009}). Thus, other approaches have attracted researchers attention to model semi-continuous data. Among those, two-part models, introduced by \cite{Duan1983} and \cite{Mullahy1986}, play an important role.
\textcolor{black}{This class of models helps handle excess of zeros and overdispersion: they involve a mixture distribution consisting in a mixing of a discrete point mass, with all mass at zero, and a discrete or continuous random variable. In particular, they are described by two-equations: a binary choice model is fitted for the probability of observing a positive-versus-zero outcome. Then, conditional on a positive outcome, an appropriate regression model is fitted for the positive outcome (see \cite{belotti2015twopm}). This statistical method introduces significant modeling flexibility by allowing the zeros and the positive values to be generated by two different processes. For these reasons, they have been deeply implemented especially in biomedical applications and in health economics as they well recover the two-step structure in the health demand process; see e.g. \cite{diehr1999methods, deb2002structure, Alfo2010, mihaylova2011review} and \cite{Maruotti2014}.}
\\
\textcolor{black}{The common structure of such models assumes that the effect of the covariates influence the mean of the conditional distribution of the response. However, in many real applications, the effect of the covariates can be different on different parts of the response distribution. In this cases it may be of interest to infer on the entire conditional distribution of the response variable using the quantile regression approach proposed in the seminal paper by \cite{koenker1978} which allows for quantile-specific inference and it is typically used for modeling non-Gaussian outcomes.}
\\
\textcolor{black}{Quantile regression methods have become widely used in literature mainly because they are suitable in all those situations where skewness, fat-tails, outliers, truncation, censoring and heteroscedasticity arise. They have been implemented in a wide range of different fields, both in a frequentist paradigm and in a Bayesian setting, spanning from medicine (see \cite{cole1992smoothing, royston1994regression, alhamzawi2012bayesian} and \cite{waldmann2018quantile}), financial and economic research (see \cite{Bassett2002, bernardi2015bayesian, petrella2018cross, laporta2018selection, tian2018quantile, bernardi2018bayesian} and \cite{petrella2019joint}) and environmental modeling; see, e.g., \cite{hendricks1992hierarchical, pandey1999comparative} and \cite{reich2011bayesian} for a discussion. For a detailed review and list of references, \cite{koenker2005} and \cite{koenker2017handbook} provide an overview of the most used quantile regression techniques in a classical setting.}
\\
\textcolor{black}{In longitudinal studies, quantile methods with random effects have been proposed in order to account for the dependence between serial observations on the same subject (see \cite{marino2015linear, alfo2017finite} and \cite{marino2018mixed}). \cite{alfo2017finite}, for example, defined a finite mixture of quantile regression models for heterogeneous data.} \textcolor{black}{Quantile regression and two-part models have also been positively considered in several studies: see for example \cite{Grilli2016, heras2018application, sauzet2019two} and \cite{biswas2020semi}. In particular, \cite{biswas2020semi} considered a semi-parametric quantile regression approach to zero-inflated and incomplete longitudinal outcomes in a repeated measurements design study.}
\\
From an inferential point of view both classical and Bayesian inferential approaches have been used in the literature to estimate the parameters and the quantiles of the models. In the frequentist setting, the inferential approach used to estimate the parameters relies on the minimization of the asymmetric loss function of \cite{koenker1978} while, in the Bayesian setting, the Asymmetric Laplace (AL) distribution has been introduced as a likelihood inferential tool (see \cite{yu2001bayesian}). The two approaches are well-justified by the relationship between the quantile loss function and the AL density: the minimization of the quantile loss function is equivalent, in terms of parameter estimates, to the maximization of the likelihood associated with the AL density. Therefore, the AL distribution could offer a convenient device to implement a likelihood based inferential approach in a quantile regression analysis.

\textcolor{black} {The main goal of the present paper is to extend the two-part quantile regression modeling framework for mixed-type outcomes to longitudinal data using a frequentist approach. In particular, we consider a mixed effect logistic regression for modeling the probability of zero-nonzero outcome, and a linear mixed quantile regression model for the continuous positive outcomes.
 Following \cite{alfo2017finite}, in order to prevent inconsistent parameter estimates due to misspecification of the random effects distribution, we adopt a non-parametric approach in which the random effect is left unspecified and approximated by using a discrete finite mixture. Within this scheme, our modeling framework reduces to a two-part finite mixture of quantile regressions where the components of the finite mixture represent clusters of units that share homogeneous values of model parameters.}
\\ 
We propose to estimate model parameters through Maximum Likelihood (ML) by using the AL distribution as a working likelihood. Specifically, estimation is carried out through the Expectation-Maximization (EM) algorithm. From a computational perspective, \textcolor{black}{we generalize the work of \cite{tian2014linear} and provide an efficient version of the EM algorithm with M-step updates in closed form using the well-known location-scale mixture representation of the AL distribution; see \cite{kozumi2011gibbs}.}
\\
\textcolor{black}{In statistical modeling, one of the main issue is the identification of the relevant variables to be considered in the model. It is in fact quite common, using real data, that a large number of predictors are concerned in the initial stage of the analysis. In this situation the researcher would be interested in determining a smaller subset that exhibits the strongest effects. Several variable selection methods have been proposed in the literature, one of them is the penalized method which is particularly useful when 
dealing with high dimensional statistical problems (see \cite{wasserman2009high} and \cite{fan2010selective}). 
 To improve estimation, to gain in parsimony and to conduct a variable selection procedure, we consider a Penalized EM (PEM) algorithm by introducing the Least Absolute Shrinkage and Selecting Operator (LASSO) $L_1$ penalty term of \cite{Tibshirani1996}.}

The relevance of our approach is also shown empirically by the analysis of a sample taken from the RAND Health Insurance Experiment (RHIE). The RHIE is one of the largest social experiments ever completed in the U.S. to study the cost sharing and its effect on service use, quality of care and health expenditures; see \cite{deb2002structure}. Healthcare data represents a striking example of semi-continuous data because they are non-negative, with substantial positive skewness, heavy tailed and, often, multi-modal, e.g. they exhibit a spike-at-zero for non-users. 
In this paper, we adopt the proposed two-part finite mixture of quantile regressions in order to investigate whether the effect of socioeconomic and household's characteristics changes with the increase in conditional health spending. \textcolor{black}{In accordance with the results of \cite{deb2002structure}, our analysis shows that the two-part model identifies two groups of users: a group of reluctant users and a second group that often uses healthcare services. In addition, the effect of the included covariates is not uniform across quantiles but it changes sign and magnitude as the quantile level varies.}\\
The rest of the paper is organized as follows. In Section \ref{sec:meth}, we introduce the two-part finite mixture quantile regression model. Section \ref{sec:est} illustrates the EM-based maximum likelihood approach, the closed form solutions and the PEM algorithm. In Section \ref{sec:app} we discuss the main empirical results while Section \ref{sec:con} concludes.

\section{Methodology}\label{sec:meth}

\subsection{Two-part quantile regression model}\label{subsec:model}
Let $y_{it}$, $i=1,..., N$, $t = 1..., T_i$ be a semi-continuous variable for unit $i$ at time $t$ and let $\textbf{b}_i = (\textbf{b}_{i0} , \textbf{b}_{i1})$ be a time-constant, individual-specific, random effects vector having distribution $f_{\bf b} (\cdot)$ with support $\mathcal{B}$ where $\mathbb{E}[\textbf{b}_i] = 0$ is used for parameter identifiability. The role of the random coefficients $\textbf{b}_{i}$ is to capture unobserved heterogeneity and within subject dependence. \textcolor{black}{In a two-part model, the probability distribution of the outcome variable $y_{it}$ can be written as}:
\begin{equation}\label{eq:hurdle}
f (y_{it}) = p_{it}^{d_{it}}\Big[ \left(1-p_{it}\right)g \big( h(y_{it})|y_{it}>0 \big) \Big]^{1-d_{it}}
\end{equation}
with $$d_{it} = \boldsymbol{1}(y_{it}=0), \quad p_{it} = \Pr(y_{it}=0) = \Pr(d_{it}=1)$$
where $d_{it}$ denotes the occurrence variable for unit $i$ at time $t$, $\boldsymbol{1}(\cdot)$ is the indicator function, $g(\cdot)$ is the density function for the positive outcome given that $y_{it}>0$ and $h(\cdot)$ is a (monotone) transformation function of $y_{it}$. 
The model is completed by defining the linear predictors for the binary and the positive parts of the model. 
\textcolor{black}{We assume that the} spike-at-zero process is governed by a binary logistic model such that:
\begin{equation}\label{eq:logit}
{\rm logit}(p_{it} \mid {\bf s}_{it}, \textbf{b}_{i0}) = {\bf s}_{it}' \boldsymbol{\gamma} + {\bf c}_{it}' \textbf{b}_{i0}
\end{equation}
where ${\bf s}_{it}=(s_{it1},\dots, s_{itm})$ is the $m$ dimensional set of explanatory variables, $\boldsymbol{\gamma}$ is the parameter vector and ${\bf c}_{it}$ is a subset of ${\bf s}_{it}$.\\
\textcolor{black}{As mentioned in the Introduction, in order to determine the effect of explanatory variables on the tails of the distribution of the outcome and make inference at an arbitrary quantile level, we model the positive outcomes using the quantile regression approach.} In the quantile regression literature, it is well established that the \textcolor{black}{likelihood approach} is based on the AL distribution of \cite{yu2001bayesian}, which gives equivalent estimates to the minimization of the loss function in \cite{koenker1978}. 
The functional form of the AL distribution \textcolor{black}{for our model} is the following:
\begin{equation}\label{eq:positive}
g(y_{it}; \mu_{it} (\tau), \sigma (\tau)) = \frac{\tau(1- \tau)}{\sigma (\tau)} \exp \Bigg\{ - \rho_\tau \left( \frac{y_{it} - \mu_{it} (\tau)}{\sigma (\tau)}\right) \Bigg\},
\end{equation}
where $\mu_{it}(\tau)$ represents the $\tau$-th quantile, with $\tau \in (0,1)$, of $y_{it}$, $\sigma (\tau) > 0$ is the scale parameter and \textcolor{black}{$\rho_\tau (\cdot)$ denotes the quantile loss function of \cite{koenker1978}:
\begin{equation}
\rho_\tau (u) = u (\tau - \boldsymbol{1}(u < 0)).
\end{equation}}
Because we are modeling positive values, to match the support of the AL density 
we consider the logarithmic transformation of the positive values of $y_{it}$. In particular, we assume that for a given $\tau$, conditionally on $\textbf{b}_{i1}$ and after log-transforming the outcome variable, $\tilde y_{it} = \log (y_{it})$, the conditional density $g( \tilde y_{it} |y_{it}>0, {\bf x}_{it}, \textbf{b}_{i1})$ in \eqref{eq:hurdle} is an AL distribution as given in \eqref{eq:positive} whose location parameter is defined by the linear model:
\begin{equation}\label{eq:lm}
\mu_{it} (\tau) = {\bf x}_{it}' \boldsymbol{\beta}(\tau) + {\bf z}_{it}' \textbf{b}_{i1}(\tau)
\end{equation}
where ${\bf z}_{it}$ is a subset of covariates of ${\bf x}_{it}$.\\
For a fixed quantile level $\tau$, \textcolor{black}{responses are assumed to be independent conditional on the random vector $\textbf{b}_i (\tau)$ and} parameter estimates can be obtained by maximizing the likelihood function of the model defined in \eqref{eq:hurdle}-\eqref{eq:lm}:
\begin{equation}\label{eq:llk1}
L({\bf \Phi_\tau}) = \prod_{i=1}^N \Bigg\{ \int_{\mathcal{B}} \prod_{t=1}^{T_i} \Big( p_{it}^{d_{it}}\left[\left(1-p_{it}\right) g_{it} \right]^{1-d_{it}} \Big) f_{\bf b} ({\bf b}_i) \textnormal{d} {\bf b}_i \Bigg\}
\end{equation} 
where \textcolor{black}{${\bf \Phi_\tau} = \{ \boldsymbol{\gamma} (\tau), \boldsymbol{\beta} (\tau), \sigma (\tau) \}$ denotes the global set of model parameters.} \textcolor{black}{The likelihood in \eqref{eq:llk1} involves a multidimensional integral over the random coefficients whose corresponding distribution $f_{\bf b} (\cdot)$ allows to explain differences in the response quantiles across individuals. 
Hence, the choice of an appropriate distribution should be data driven and resistant to misspecification \citep{marino2015linear}. In the next Section, we will discuss how we may avoid evaluating the integral in \eqref{eq:llk1} for ML estimation.}

\subsection{Finite mixture of quantile regressions}\label{sub:mix}
\textcolor{black}{In the literature, typically the Gaussian distribution is a convenient choice for $f_{\bf b} (\cdot)$ from a computational point of view. In this case, we may approximate the integral in \eqref{eq:llk1} using Gaussian quadrature or adaptive Gaussian quadrature schemes (see \cite{winkelmann2004health} and \cite{rabe2005maximum}). A disadvantage of such approaches lies in the required computational effort, which is exponentially increasing with the dimension of the random parameter vector. For these reasons, potential alternatives availed themselves of simulation methods such as Monte Carlo and simulated ML approaches. However, for samples of finite size and short individual sequences, these methods may not provide a good approximation of the true mixing distribution \citep{alfo2017finite}. As a robust alternative to the Gaussian choice, a Symmetric Laplace or a multivariate Student T random variable have been considered by \cite{Geraci2014} and \cite{farcomeni2015longitudinal}. However, a parametric assumption on the distribution of the random coefficients could be rather restrictive and misspecification of the mixing distribution could lead to biased parameter estimates (see \cite{Alfo2010}). In view of these considerations, in this work we exploit the approach based on the nonparametric maximum likelihood (NPML) estimation of \cite{laird1978nonparametric}.} \textcolor{black}{Instead of specifying parametrically the distribution $f_{\bf b} (\cdot)$ we approximate it
by using a discrete distribution on $G < N$ locations $\textbf{b}_{k} (\tau) = (\textbf{b}_{0k}(\tau), \textbf{b}_{1k}(\tau))$ i.e. ${\bf b}_i(\tau) \sim \sum_{k=1}^G \pi_k(\tau) \delta_{{\bf b}_k(\tau)}$ where the probability $\pi_k(\tau)$ is defined by $\pi_k(\tau) = \textnormal{Pr}({\bf b}_i (\tau) = {\bf b}_k (\tau) )$, $i = 1, . . . , N$ for $k = 1, . . . , G$ and  $\delta_{{\bf b}_k(\tau)}$ is a one-point distribution putting a unit mass at ${\bf b}_k (\tau)$. }
\textcolor{black}{The proposed approach can be thought as an approximation of a fully parametric framework as the discrete support approximates a possibly continuous distribution for the random coefficients.} \textcolor{black}{To ease the notation, hereinafter we omit the quantile level $\tau$ but all parameters are allowed to depend on it.} In this setting, the likelihood in \eqref{eq:llk1} reduces to:
\begin{equation}\label{eq:llk2}
L({\bf \Phi_\tau}) = \prod_{i=1}^N \Bigg\{ \sum_{k=1}^G \prod_{t=1}^{T_i} \Big( p_{itk}^{d_{it}}\left[\left(1-p_{itk}\right) g_{itk} \right]^{1-d_{it}} \Big) \pi_k \Bigg\},
\end{equation}
\textcolor{black}{where ${\bf \Phi_\tau} = \{ \boldsymbol{\gamma}, \boldsymbol{\beta}, \sigma, {\bf b}_1, . . . , {\bf b}_G, \pi_1, . . . , \pi_G \}$ is the parameter vector.\\
The likelihood in \eqref{eq:llk2} is similar to the likelihood of a finite mixture of quantile regressions with $G$ clusters. More specifically, in the $k$-th cluster the spike-at-zero process is governed by the binary logistic model, ${\rm logit}(p_{itk} \mid {\bf s}_{it}, \textbf{b}_{0k}) = {\bf s}_{it}' \boldsymbol{\gamma} + {\bf c}_{it}' \textbf{b}_{0k}$; meanwhile the positive outcomes process is regulated by a linear mixed quantile with AL density in \eqref{eq:positive} having location parameter given by $\mu_{itk} = {\bf x}_{it}' \boldsymbol{\beta} + {\bf z}_{it}' \textbf{b}_{1k}$.}\\ 
Therefore, our modeling framework reduces to a finite bivariate mixture model for each quantile level where heterogeneity sources that influence the binary decision process, are assumed to influence also the distribution of the positive outcomes through the latent structure defined by discrete multivariate random effects.

\section{Estimation}\label{sec:est}
In this Section, we propose a maximum likelihood approach based on the EM algorithm to estimate the parameters of the methodology illustrated in Section \ref{sec:meth}. Given the finite mixture representation in \eqref{eq:llk2}, each unit $i$ can be conceptualized as drawn from one of $G$ distinct groups: we denote with $w_{ik}$ the indicator variable that is equal to $1$ if the $i$-th unit belongs to the $k$-th component of the finite mixture, and 0 otherwise. The EM algorithm treats as missing data the component membership $w_{ik}$. 
Thus, the log-likelihood for the complete data has the following form:
\begin{fleqn}[\parindent]
\begin{align}
\ell_c ( {\bf \Phi_\tau}) & = \sum_{i=1}^N \sum_{k=1}^G w_{ik} \Bigg\{ \sum_{t=1}^{T_i} \log \Bigg( p_{itk}^{d_{it}}\left[\left(1-p_{itk}\right) g_{itk} \right]^{1-d_{it}} \Bigg) + \log (\pi_k) \Bigg\}. \label{eq:cdl}
\end{align}
\end{fleqn}
In the E-step, the presence of the unobserved group-indicator $w_{ik}$ is handled by taking the conditional expectation of $w_{ik}$ given the observed data and the parameter estimates at the $r$-th iteration $\hat{{\bf \Phi}}^{(r)}_\tau = \{ \hat{\boldsymbol{\gamma}}^{(r)}, \hat{\boldsymbol{\beta}}^{(r)}, \hat{\sigma}^{(r)}, \hat{{\bf b}}_1^{(r)}, . . . , \hat{{\bf b}}_G^{(r)}, \hat{\pi}_1^{(r)}, . . . , \hat{\pi}_G^{(r)}  \}$. At the $(r + 1)$-th iteration of the algorithm, we replace $w_{ik}$ by its conditional expectation $\hat{w}^{(r+1)}_{ik}$ using the following update equation:
\begin{equation}\label{eq:weights}
\hat{w}^{(r+1)}_{ik} = \mathbb{E} [ w_{ik} | y_{it}, {\bf s}_{it}, {\bf x}_{it}, {\bf \hat{\Phi}}^{(r)}_\tau] =  \frac{\prod_{t=1}^{T_i} f^{(r)}_{itk} \hat{\pi}_k^{(r)}}{\sum_{l=1}^G \prod_{t=1}^{T_i} f^{(r)}_{itl} \hat{\pi}_l^{(r)}},
\end{equation}
where $f^{(r)}_{itk} = p_{itk}(\hat{{\bf \Phi}}^{(r)}_\tau) ^{d_{it}} \left[\left(1-p_{itk} (\hat{{\bf \Phi}}^{(r)}_\tau) \right) g_{itk} (\hat{{\bf \Phi}}^{(r)}_\tau) \right]^{1-d_{it}}$. 
Conditionally on the posterior probabilities $\hat{w}^{(r+1)}_{ik}$ in \eqref{eq:weights}, the M-step solutions are generally updated by maximizing $\mathbb{E}[ \ell_c ( {\bf \Phi_\tau}) \mid y_{it}, {\bf s}_{it}, {\bf x}_{it}, \hat{{\bf \Phi}}^{(r)}_\tau]$ with respect to ${\bf \Phi_\tau}$ using numerical optimization techniques. 
The E- and M-steps are alternated until convergence, \textcolor{black}{that is when the difference between the likelihood function evaluated at two consecutive iterations is smaller than a predetermined threshold. In this paper, we set this convergence criterion equal to $10^{−5}$.} To avoid convergence to local maxima, for each value of $G$, we initialize model parameters using a multi-start strategy: we considered 20 different starting points and retained the solution corresponding to the maximum likelihood value.\\
\textcolor{black}{A considerable disadvantage of this procedure affects the M-step which could require a high computational effort and it could be very time-consuming especially when the set of explanatory variables is large. Therefore, in the next Section we develop a more efficient approach for estimating two-part quantile regression models, and to obtain a closed form of the ML estimator.}



\subsection{Closed form EM algorithm solutions}\label{sub:closed}
\textcolor{black}{This Section develops an efficient estimation method for the two-part quantile regression problem based on the EM algorithm presented in Section \ref{sec:est}. We reduce the computational burden of the algorithm compared to direct maximization of the likelihood and extend the work of \cite{tian2014linear} to two-part mixture models. In particular, we obtain iteratively closed form expressions for the unknown parameters vector $\boldsymbol{\beta}$ and the coefficient vectors ${\bf b}_{1k}$ for $k=1, \dots, G$.}\\
We use the location-scale mixture representation of the AL density considered in \cite{kozumi2011gibbs} to specify the positive outcome process of the model in \eqref{eq:lm} as the following hierarchical model:
\begin{equation}\label{eq:hier}
\tilde y_{it} \mid ( y_{it}>0, {\bf x}_{it}, \textbf{b}_{1k}, v_{it}) \sim N(\mu_{itk} + \theta v_{it}, \rho^2 \sigma v_{it}), \quad v_{it} \sim \textnormal{Exp} (\frac{1}{\sigma}),
\end{equation}
where $\mu_{itk} = {\bf x}_{it}' \boldsymbol{\beta} + {\bf z}_{it}' \textbf{b}_{1k}$, $\theta = \frac{1-2\tau}{\tau (1-\tau)}$ and $\rho^2 = \frac{2}{\tau (1-\tau)}$. The constraints imposed on $\theta$ and $\rho$ must be satisfied in order to guarantee that $\mu_{itk}$ coincides with the $\tau$-th conditional quantile of $\tilde y_{it} \mid (y_{it}>0, {\bf x}_{it}, \textbf{b}_{1k})$.\\
Due to the independence of the $v_{it}$'s, one can obtain that the conditional distribution of $v_{it}$ is a Generalized Inverse Gaussian (GIG) distribution (see \citealt{tian2014linear}, Section 2.3), namely:
\begin{equation}\label{eq:GIG}
f(v_{it} | \tilde y_{it}, y_{it}>0, {\bf x}_{it}, \textbf{b}_{1k}) \sim \textnormal{GIG} \Bigg( \frac{1}{2}, \frac{(\tilde y_{it} - \mu_{itk})^2}{\rho^2 \sigma} , \frac{2\rho^2 + \theta^2}{\rho^2 \sigma} \Bigg).
\end{equation}
In addition, from \eqref{eq:hier} the joint density of $\tilde y_{it}$ and $v_{it}$ is:
\begin{equation}\label{eq:ytildev}
f(\tilde y_{it}, v_{it} | y_{it}>0, {\bf x}_{it}, \textbf{b}_{1k}) = \frac{1}{ \sqrt{2 \pi \sigma v_{it}} \sigma \rho} \exp \bigg( - \frac{(\tilde y_{it} - \mu_{itk} - \theta v_{it})^2}{2 \rho^2 \sigma v_{it}} -\frac{v_{it}}{\sigma} \bigg).
\end{equation}
The underlying idea to obtain the updated parameter estimates of $\boldsymbol{\beta}$ and ${\bf b}_{1k}$ for $k=1, \dots, G$ is to consider $v_{it}$ as an additional latent variable. According to \eqref{eq:ytildev}, after omitting terms which of are not dependent on $\boldsymbol{\beta}$ and ${\bf b}_{1k}$, the complete data log-likelihood function is proportional to:
\begin{fleqn}
\begin{align}
& \ell_c ( \boldsymbol{\beta}, {\bf b}_{11}, ..., {\bf b}_{1G}) \propto \frac{1}{2} \sum_{i=1}^N \sum_{k=1}^G \sum_{t=1}^{T_i}  w_{ik} (1-d_{it}) v_{it}^{-1} (\tilde y_{it} - {\bf x}_{it}' \boldsymbol{\beta} - {\bf z}_{it}' \textbf{b}_{1k})^2 \label{eq:cdl2} \\
& - \theta \sum_{i=1}^N \sum_{k=1}^G \sum_{t=1}^{T_i}  w_{ik} (1-d_{it}) (\tilde y_{it} - {\bf x}_{it}' \boldsymbol{\beta} - {\bf z}_{it}' \textbf{b}_{1k}).
\end{align}
\end{fleqn}
In the E-step, the conditional expectation of the complete log-likelihood function, given the observed data and the current parameter estimates at the $r$-th iteration, $\hat{{\bf \Phi}}^{(r)}_\tau$, is given by:
\begin{fleqn}
\begin{align}
\mathbb{E}[ \ell_c ( \boldsymbol{\beta}, {\bf b}_{11}, ..., {\bf b}_{1G}) \mid y_{it}, {\bf s}_{it}, {\bf x}_{it}, \hat{{\bf \Phi}}^{(r)}_\tau]
\end{align}
\end{fleqn}
\begin{fleqn}
\begin{align}
& \propto \frac{1}{2} \sum_{i=1}^N \sum_{k=1}^G \sum_{t=1}^{T_i} \hat{w}_{ik}^{(r+1)} (1-d_{it}) \hat{v}_{it}^{(r+1)} (\tilde y_{it} - {\bf x}_{it}' \boldsymbol{\beta} - {\bf z}_{it}' \textbf{b}_{1k})^2 \label{eq:ecdl21} \\ 
& - \theta \sum_{i=1}^N \sum_{k=1}^G \sum_{t=1}^{T_i}  \hat{w}_{ik}^{(r+1)} (1-d_{it}) (\tilde y_{it} - {\bf x}_{it}' \boldsymbol{\beta} - {\bf z}_{it}' {\bf b}_{1k}), \label{eq:ecdl22}
\end{align}
\end{fleqn}
where $\hat{w}_{ik}^{(r+1)}$ has been defined in \eqref{eq:weights} and $\hat{v}_{it}^{(r+1)} = \mathbb{E}[v_{it}^{-1} \mid y_{it}, {\bf s}_{it}, {\bf x}_{it}, \hat{{\bf \Phi}}^{(r)}_\tau]$. To compute $\hat{v}_{it}^{(r+1)}$, we can exploit the moment properties of the GIG distribution in \eqref{eq:GIG}. Hence, we have that:
\begin{equation}\label{eq:mominv}
\hat{v}_{it}^{(r+1)} = \mathbb{E}[v_{it}^{-1} \mid y_{it}, {\bf s}_{it}, {\bf x}_{it}, \hat{{\bf \Phi}}^{(r)}_\tau] = \frac{\sqrt{\theta^2 + 2\rho^2}}{\mid \tilde y_{it} - {\bf x}_{it}' \boldsymbol{\hat{\beta}}^{(r)} - {\bf z}_{it}' {\bf \hat{b}}_{1k}^{(r)} \mid}.
\end{equation}
In the M-step, we determine the update expressions by setting to zero the derivative of \eqref{eq:ecdl21}-\eqref{eq:ecdl22} with respect to $\boldsymbol{\beta}$ and ${\bf b}_{11}, ..., {\bf b}_{1G}$ and solve the corresponding M-step equations. 
In conclusion, we obtain the following update expressions:
\begin{equation}\label{eq:beta}
\scriptsize
\boldsymbol{\hat{\beta}}^{(r+1)} = \bigg( \sum_{i=1}^N \sum_{k=1}^G \sum_{t=1}^{T_i} \hat{w}_{ik}^{(r+1)} (1-d_{it}) \hat{v}_{it}^{(r+1)} {\bf x}_{it} {\bf x}_{it}' \bigg)^{-1} \bigg( \sum_{i=1}^N \sum_{k=1}^G \sum_{t=1}^{T_i} \hat{w}_{ik}^{(r+1)} (1-d_{it}) \big( \hat{v}_{it}^{(r+1)} {\bf x}_{it} (\tilde y_{it} - {\bf z}_{it}' {\bf \hat{b}}_{1k}^{(r)}) - \theta {\bf x}_{it} \big) \bigg)
\end{equation}
and
\begin{equation}\label{eq:bpos}
\scriptsize
\hat{{\bf b}}_{1k}^{(r+1)} = \bigg( \sum_{i=1}^N \sum_{t=1}^{T_i} \hat{w}_{ik}^{(r+1)} (1-d_{it}) \hat{v}_{it}^{(r+1)} {\bf z}_{it} {\bf z}_{it}' \bigg)^{-1} \bigg( \sum_{i=1}^N \sum_{t=1}^{T_i} \hat{w}_{ik}^{(r+1)} (1-d_{it}) \big( \hat{v}_{it}^{(r+1)} {\bf z}_{it} (\tilde y_{it} - {\bf x}_{it}' \boldsymbol{\hat{\beta}}^{(r)}) - \theta {\bf z}_{it} \big) \bigg).
\end{equation}
Equations \eqref{eq:beta} and \eqref{eq:bpos} essentially, are equivalent to modified weighted least square estimator expressions where the quantile level $\tau$ is used to modify both the weights, $\hat{w}_{ik}^{(r)} \hat{v}_{it}^{(r)}$, and the response variable through $\hat{v}_{it}^{(r)}$ and $\theta$.

\subsection{Variable selection and the Penalized EM algorithm}\label{sub:penest}
\textcolor{black}{When dealing} with high-dimensional problems, \textcolor{black}{it may be of interest to reduce the set of explanatory variables using penalized regression methods which allow for sparse modeling and enhance model interpretability.}
In this Section, we introduce a penalized version of the EM algorithm described in Section \ref{sec:est}. In particular, we use the PEM algorithm, originally proposed by \cite{green1990use}, which leaves the E-step unchanged while it modifies the M-step by introducing a penalty function to achieve simultaneous shrinkage and/or variable selection. Here we consider the LASSO penalty term put forward by \cite{Tibshirani1996} to shrink the coefficients of the positive outcomes, i.e. the vector $\boldsymbol{\beta}$, and
 simultaneously select a smaller subset of variables that exhibits the strongest effects. For a chosen quantile level $\tau$ and number of latent classes $G$, the penalized log-likelihood for the complete data has the following form:
\begin{equation}\label{eq:pencdl}
\ell_{pen} ({\bf \Phi_\tau} | \lambda_\tau) = \ell_c({\bf \Phi_\tau}) - \lambda_\tau J(\boldsymbol{\beta}),	
\end{equation}
where $\ell_c({\bf \Phi_\tau})$ has been defined in \eqref{eq:cdl}, $J(\boldsymbol{\beta}) = \mid \mid \boldsymbol{\beta} \mid \mid_1$ is the LASSO penalty function and $\lambda_\tau$ is a tuning parameter that regulates the strength of the penalization assigned to the coefficients of the model. The optimal value of $\lambda_\tau$ is selected via 10-fold cross-validation which allows us to consider $\lambda_\tau$ as a data-driven parameter. 

\section{Application}\label{sec:app}


\subsection{Data description}
\begin{figure}[h]
\center
\includegraphics[width=1\linewidth, height=6.5cm, keepaspectratio]{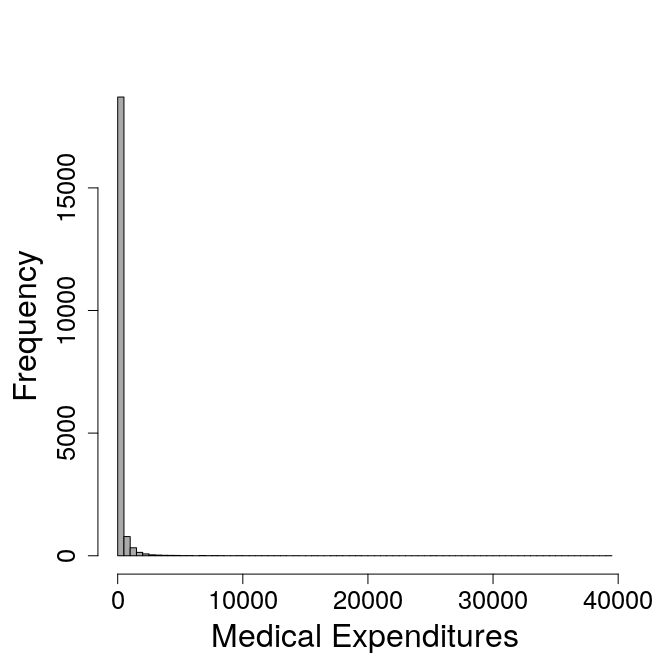}
\includegraphics[width=1\linewidth, height=6.5cm, keepaspectratio]{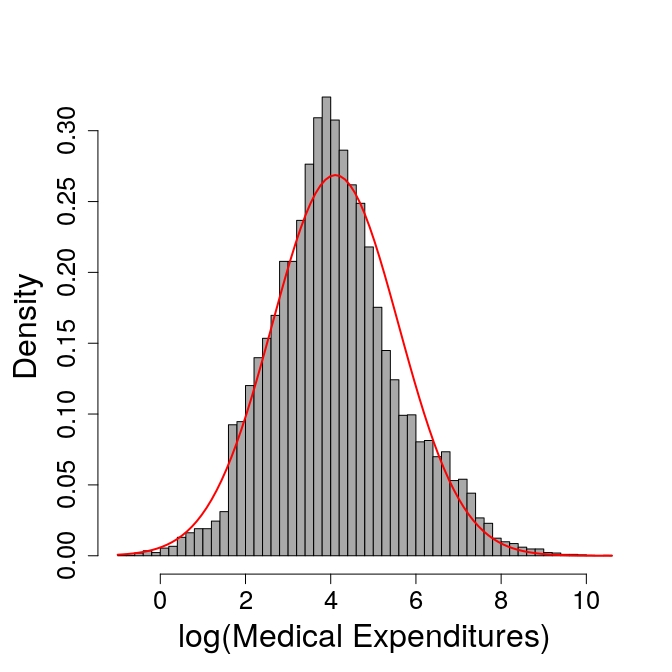}
\caption{Medical expenditure distribution: marginal distribution (left) and distribution of the logarithm of positive values (right).}\label{fig:data}
\end{figure}

As stated in the Introduction, in this Section we present the application of the proposed methodology to the well-known RHIE dataset. \textcolor{black}{These data have already been discussed by \cite{deb2002structure, Duan1983} and \cite{manning1987health}.} The RHIE is the most important health insurance study ever conducted to assess how medical care costs affect a patient’s use of health services and quality and it is widely regarded as the basis of the most reliable estimates of price sensitivity of demand for medical services. 
The experiment, conducted by the RAND Corporation from 1974 to 1982, collects data from about 8000 enrollees in 2823 families, from six sites across the US. Each family was enrolled in one of fourteen different HIS insurance plans for either 3 or 5 years.\\
Our aim is to understand how available covariates influence healthcare decisions spending in U.S. families at different quantile levels of interest. We consider one measure of utilization: the total spending on health services (MED) defined as the sum of outpatient, inpatient, drugs, supplies and psychotherapy expenses expressed in U.S. dollars. The considered covariates are the same of the work of \cite{deb2002structure}: 
they include personal characteristics such as sex (FEMALE), age (XAGE), race (BLACK) and education level (EDUCDEC). There are also socio-economic variables of the household such as income (LINC), number of components (LFAM) and the presence of a person aged 18 or less (CHILD). An interaction term between the gender and the presence of a child (FEMCHILD) is also included. Quantitative indicators of health condition are measured via an index of chronic conditions (DISEA) and through the existence of a physical limitation (PHYSLM) while the binary coding of self-rated health status (HLTHG, HLTHF, HLTHP) controls for variations in perceived healthcare conditions. The summary statistics 
are reported in Tables \ref{tab:y} and \ref{tab:X}. \textcolor{black}{By looking at Table \ref{tab:y}, the proportion of zero expenditures is significant: 22\% did not have any medical expenditure. Meanwhile, the right tail of the distribution is very long with the maximum expense being 39182.02 U.S. dollars which indicates the presence of overdispersion and potential outliers in the data. In addition, the dependent variable is severely skewed and presents a high kurtosis. From a graphical point of view, Figure \ref{fig:data} shows that the data is characterized by a substantial zero-mass health expenditure (left graph): zero expenditure indicates no utilization, and it may reflect population' reluctance to spend on health care treatments. By looking at the right plot, also the logarithmic transformation of positive values does not follow a Gaussian distribution. Therefore, a two-part quantile regression model would be appropriate for these data. In particular, the binary decision process would estimate the association between the covariates and the probability of having any health care expenditure. Through the positive part of the model instead, our goal is to see whether the covariates have a different impact on health care expenditures at different quantiles. Indeed, by doing so, we can single out the impact of health determinants across extreme quantiles and non extreme quantiles reflecting the association with low-intensity primary health care services, at the 10-th and 25-th percentiles, and that with high-intensity such as the highly intensive and expensive technology care, at the 75-th and 90-th percentiles.}

\begin{table}[h]
\centering
 \smallskip 
 \resizebox{1.0\columnwidth}{!}{%
\begin{tabular}{lcccccccccccc}
  \hline
Variable & Mean & S.D. & Skewness & Kurtosis & No Exp. (\%) & Maximum &\multicolumn{5}{c}{$\tau$-th quantile} \\\cmidrule(r){8-12}
  & & & & & & & 0.1 & 0.25 & 0.5 & 0.75 & 0.90\\  
 \hline
MED & 171.568 & 698.201 & 20.189 & 734.007 & 22.055 & 39182.016 & 0.000 & 5.503 & 35.378 & 104.541 & 341.201 \\ 
   \hline
\end{tabular}}
\caption{Summary statistics of healthcare expenditures.}\label{tab:y}
\end{table}

\begin{table}[h]
\centering
 \smallskip 
 \resizebox{0.8\columnwidth}{!}{%
\begin{tabular}{llrrr}
  \hline
 Variable & Description & Mean & S.D. \\ 
  \hline
LOGC & $\log$(coinsurance + 1), $0 \leq$ coinsurance $\leq 100$ & 2.384 & 2.042 \\
LFAM & $\log$(family size) & 1.248 & 0.539 \\ 
  LINC & $\log$(family income) & 8.708 & 1.228 \\ 
  XAGE & Age in years & 25.718 & 16.768 \\ 
  FEMALE & If person is female: 1 & 0.517 & 0.500 \\ 
  CHILD & If age is less than 18: 1 & 0.401 & 0.490 \\ 
  FEMCHILD & FEMALE ∗ CHILD & 0.194 & 0.395 \\ 
  BLACK & If race of household head is black: 1 & 0.184 & 0.387 \\ 
  EDUCDEC & Education of the household head in years & 11.966 & 2.806 \\ 
  PHYSLM & If the person has a physical limitation: 1 & 0.118 & 0.323 \\ 
  DISEA & Index of chronic diseases & 11.244 & 6.742 \\ 
  HLTHG & If self-rated health is good: 1$^\dagger$
 & 0.362 & 0.481 \\ 
  HLTHF & If self-rated health is fair: 1$^\dagger$
 & 0.077 & 0.267 \\ 
  HLTHP & If self-rated health is poor: 1$^\dagger$
 & 0.015 & 0.121 \\ 
  MHI & Mental Health Index & 76.554 & 12.503 \\
   \hline
\end{tabular}}
\caption{Covariates description and summary statistics. $^\dagger$ indicates that the baseline individual is in excellent self-rated health.}\label{tab:X}
\end{table}

\subsection{Marginal inferences from the two-part mixture quantile regression model}
In this Section we present the results of the application to the RHIE data. We estimated the proposed \textcolor{black}{LASSO penalized} two-part finite mixture quantile regression at different quantile levels of interest $ \tau = (0.1, 0.25, 0.5, 0.75, 0.9)$, for a varying number of mixture components $G = (1,\dots, 6)$. We used the same covariates for both the binary process and the positive part of the model and we assumed that random intercepts are sufficient to capture heterogeneity between subjects even tough 
the generalization to multiple random effects is computationally inexpensive and straightforward thanks to the discrete mixture structure. \textcolor{black}{Because the number of components $G$ is unknown a-priori, we select it according to the Bayesian Information Criterion (BIC)\footnote{$\textnormal{BIC} = -2 \log (L) + \nu \log(N)$, where $\nu$ denotes the number of free model parameters and $N$ is the number of individuals.} (\cite{schwarz1978estimating}). Standard errors are obtained using a parametric bootstrap approach: we refitted the model to 250 bootstrap samples simulated from the estimated model parameters, and approximated the standard error of each parameter with its corresponding standard deviation computed on bootstrap samples.} All computations have been implemented in the \verb!R! software environment (version 3.5.2) with \verb!C++! object-oriented programming.\\
Table \ref{tab:pen} reports the estimated penalized model coefficients and standard errors (in parentheses) for the binary (Panel A) and positive (Panel B) part of the model. Parameter estimates are displayed in boldface when significant at the standard 5\% level. Panels C and D show the estimated masses of the bivariate discrete distribution of the random effects and the model fit summary at each quantile level, respectively.\\
\textcolor{black}{Consistently with the findings of \cite{deb2002structure, Duan1983} and \cite{bago2006latent}}, the selected number of mixture components $G$ is two for all investigated quantiles. Essentially, the two-part model demarcates two distinct subpopulations: a group of reluctant users who rarely accesses health services and a group of users that uses health services with high frequency. This claim is confirmed by the estimated mixing probabilities $(\pi_1, \pi_2)$ and locations $\textbf{b}_{1}, \textbf{b}_{2}$ which can be interpreted as high and low users. Group separation is all the more apparent as the quantile level increases with estimated masses $(0.328, 0.672)$ at the 90-th quantile.\\
By looking at the binary process estimates in Table \ref{tab:pen} (Panel A), one can see that all coefficients are statistically different from zero at the 5\% significance level except LFAM, MHI and HLTHG and HLTHF at low and high quantile levels, respectively. The marginal inferences from the binary part of the two-part model suggest that the most important determinants of people's willingness to spend on health services are related to socio-cultural and economic factors: income, age, sex, race ethnicity and education level have the expected signs and significance. The presence of a coinsurance health plan acts as a protection factor reducing the probability of consuming health services. Ageing, presumably, involves complications such as heart diseases, cancer, diabetes and many others that require (obligatory) hospitalization. The analysis points out also profound economic and ethnicity inequalities. \\
It is worth noting that the coefficient estimates vary as the quantile level varies, e.g. they generally show an upward or downward trend as the quantile increases. This may be due to the presence of correlation between the binary and the positive processes which is channeled via the latent structure defined by the discrete bivariate random effects.\\
Moving onto the analysis of the positive outcomes (Table \ref{tab:pen}, Panel B), our results are generally consistent with those of \cite{deb2002structure}. The first thing one can notice is that not all variables have the same effect on healthcare consumption, i.e. they change in sign and magnitude and some of them have been set to zero at specific quantile levels. \textcolor{black}{This highlights the importance of considering a quantile regression approach because such effects could not be detected by the classical mean regression.}
 As previously described, the presence of health insurance seems to act as a safety ledge: insured people spend less. This may reflect the fact that health care is largely provided by private hospitals and clinics. The results indicate also that individuals in larger families are more likely to be in the low-use group especially up to the 75-th quantile. As expected, the economic situation appears to be one of the main barriers that influence healthcare access for low income households. 
In relation to the gender, females have an higher spending pattern up to the 75-th percentiles which may be caused by hospitalization being parturition or its complications.\\
 Other covariates also appear to have some explanatory power, notably XAGE, CHILD, FEMCHILD and BLACK. The elderly undergoes frequent checks and require more health services due to ageing phenomena to which they may be exposed to. 
The interaction term, FEMCHILD, and BLACK are negatively associated with the response indicating that, in the latter case, racial differences in the amount spent on medical care are persistent and confirms that among individuals who spend on care, blacks have lower expenditures than whites. Moreover, highly literate and educated individuals tend to follow a healthy lifestyle, be less reluctant to go through a visit to a physician, and thus spend more to preserve and improve their health. Variables controlling for the health status of the individual such as PHYSLM and DISEA are positively associated with healthcare expenses: supposedly, disabilities, physical limitations and diagnosis of chronic illnesses require highly intensive, technology care or newer, more costly treatments. Also, self-perceived health condition indicators are all positively associated with health expenses with a much steeper slope at the lower percentiles meanwhile, their impact seems to be negligible above the 75-th quantile.\\
\textcolor{black}{Moreover, it is possible to see that the impact of several variables varies across quantiles: the effect of LOGC and BLACK is not uniform as the quantile increases but is worse on the left tail of the distribution of health expenditures.}
Also, variables such as LINC, FEMALE, CHILD and FEMCHILD exhibit nonlinear effects: in particular, CHILD and FEMCHILD have a U-shaped and an inverted U-shape effect, respectively. EDUCDEC changes sign and magnitude across the quantile levels as well. As regards the estimated random intercepts, $\textbf{b}_{1k}$, we notice that the estimates increase with $\tau$ and this is consistent with increasing values of healthcare spending.\\
\textcolor{black}{We conclude the analysis comparing our methodology with the Linear Quantile Mixed Model (LQMM) of \cite{Geraci2014}. In particular, we fit a LQMM only on the positive values of the dependent variable under the assumption that the distribution of the random intercepts is a bivariate Gaussian while ignoring both the correlation between the binary and positive process and the sparsity induced through the LASSO regularization. 
By looking at the parameter estimates in Table \ref{tab:unpen}, we notice slight differences with respect to the results obtained using our methodology at low quantiles while, as we move from the left to the right tail of the response distribution, there are considerable differences. Such discrepancies in the two models might be traced back to the fact that the distribution of the random effects is not Gaussian. Indeed, the semiparametric mixture approach and the LQMM perform equivalently only if the random coefficients are Gaussianly distributed otherwise, the semiparametric mixture performs better being more flexible and able to accommodate departure from the Gaussianity assumption (see \cite{alfo2017finite}).}

%

\begin{table}
\centering
 \smallskip 
 \resizebox{0.85\columnwidth}{!}{
 \setlength\tabcolsep{17pt}
\begin{tabular}{l c c c c c }
\hline
Covariate &\multicolumn{5}{c}{$\tau$-th quantile} \\\cmidrule(r){2-6}
  & 10 & 25 & 50 & 75 & 90 \\ 
\hline
Panel A: Binary process ($\mathbb{P}\textnormal{r}(y_{it} = 0)$)\\ 

LOGC     & $\mathbf{0.413} \; (0.022)$  & $\mathbf{0.414} \; (0.025)$  & $\mathbf{0.406} \; (0.026)$  & $\mathbf{0.379} \; (0.022)$  & $\mathbf{0.377} \; (0.019)$  \\
LFAM     & $-0.039 \; (0.025)$          & $-0.016 \; (0.026)$          & $-0.021 \; (0.027)$          & $0.014 \; (0.024)$           & $-0.017 \; (0.022)$          \\
LINC     & $\mathbf{-0.190} \; (0.022)$ & $\mathbf{-0.196} \; (0.024)$ & $\mathbf{-0.186} \; (0.025)$ & $\mathbf{-0.175} \; (0.022)$ & $\mathbf{-0.097} \; (0.019)$ \\
XAGE     & $\mathbf{-0.101} \; (0.037)$ & $-0.066 \; (0.039)$          & $-0.059 \; (0.041)$          & $\mathbf{-0.082} \; (0.038)$ & $\mathbf{-0.120} \; (0.034)$ \\
FEMALE   & $\mathbf{-0.841} \; (0.061)$ & $\mathbf{-0.974} \; (0.067)$ & $\mathbf{-0.972} \; (0.067)$ & $\mathbf{-0.860} \; (0.059)$ & $\mathbf{-0.689} \; (0.054)$ \\
CHILD    & $\mathbf{-0.450} \; (0.088)$ & $\mathbf{-0.465} \; (0.094)$ & $\mathbf{-0.426} \; (0.096)$ & $\mathbf{-0.511} \; (0.087)$ & $\mathbf{-0.432} \; (0.080)$ \\
FEMCHILD & $\mathbf{0.882} \; (0.087)$  & $\mathbf{0.919} \; (0.095)$  & $\mathbf{0.921} \; (0.094)$  & $\mathbf{0.873} \; (0.088)$  & $\mathbf{0.695} \; (0.074)$  \\
BLACK    & $\mathbf{1.422} \; (0.058)$  & $\mathbf{1.351} \; (0.059)$  & $\mathbf{1.443} \; (0.066)$  & $\mathbf{1.229} \; (0.049)$  & $\mathbf{1.134} \; (0.045)$  \\
EDUCDEC  & $\mathbf{-0.160} \; (0.023)$ & $\mathbf{-0.168} \; (0.024)$ & $\mathbf{-0.192} \; (0.025)$ & $\mathbf{-0.151} \; (0.021)$ & $\mathbf{-0.163} \; (0.020)$ \\
PHYSLM   & $\mathbf{-0.543} \; (0.078)$ & $\mathbf{-0.725} \; (0.082)$ & $\mathbf{-0.720} \; (0.088)$ & $\mathbf{-0.590} \; (0.078)$ & $\mathbf{-0.500} \; (0.073)$ \\
DISEA    & $\mathbf{-0.287} \; (0.028)$ & $\mathbf{-0.278} \; (0.028)$ & $\mathbf{-0.276} \; (0.029)$ & $\mathbf{-0.237} \; (0.027)$ & $\mathbf{-0.246} \; (0.024)$ \\
HLTHG    & $\mathbf{0.123} \; (0.048)$  & $0.066 \; (0.053)$           & $0.020 \; (0.054)$           & $\mathbf{0.124} \; (0.047)$  & $\mathbf{0.117} \; (0.044)$  \\
HLTHF    & $\mathbf{-0.207} \; (0.091)$ & $-0.178 \; (0.098)$          & $\mathbf{-0.272} \; (0.108)$ & $\mathbf{-0.271} \; (0.089)$ & $-0.131 \; (0.083)$          \\
HLTHP    & $\mathbf{-1.026} \; (0.246)$ & $\mathbf{-0.996} \; (0.258)$ & $\mathbf{-1.011} \; (0.259)$ & $\mathbf{-0.903} \; (0.231)$ & $\mathbf{-0.635} \; (0.209)$ \\
MHI      & $-0.016 \; (0.023)$          & $-0.037 \; (0.025)$          & $\mathbf{-0.057} \; (0.026)$ & $-0.026 \; (0.022)$          & $-0.030 \; (0.021)$          \\
$\textbf{b}_{01}$   & $\mathbf{-2.671} \; (0.072)$ & $\mathbf{-2.781} \; (0.081)$ & $\mathbf{-2.897} \; (0.086)$ & $\mathbf{-2.803} \; (0.084)$ & $\mathbf{-2.193} \; (0.066)$ \\
$\textbf{b}_{02}$   & $\mathbf{-0.476} \; (0.052)$ & $\mathbf{-0.326} \; (0.056)$ & $\mathbf{-0.299} \; (0.057)$ & $\mathbf{-0.686} \; (0.053)$ & $\mathbf{-0.904} \; (0.049)$ \\
\hline

Panel B: Positive process\\
LOGC     & $-0.196$ & $-0.192$ & $-0.142$ & $-0.091$ & $-0.081$ \\
LFAM     & $-0.075$ & $-0.098$ & $-0.057$ & $-0.106$ & $-$  \\
LINC     & $0.103$  & $0.088$  & $0.068$  & $0.122$  & $-$  \\
XAGE     & $0.219$  & $0.194$  & $0.165$  & $0.154$  & $0.359$  \\
FEMALE   & $0.287$  & $0.351$  & $0.270$  & $0.400$  & $-$  \\
CHILD    & $0.096$  & $-$  & $-0.204$ & $-0.097$ & $-0.001$ \\
FEMCHILD & $-0.299$ & $-0.259$ & $-0.230$ & $-0.461$ & $-0.006$ \\
BLACK    & $-0.547$ & $-0.379$ & $-0.436$ & $-0.243$ & $-$  \\
EDUCDEC  & $0.054$  & $0.039$  & $0.044$  & $-0.017$ & $0.012$  \\
PHYSLM   & $0.189$  & $0.404$  & $0.371$  & $0.437$  & $0.281$  \\
DISEA    & $0.186$  & $0.158$  & $0.139$  & $0.073$  & $0.072$  \\
HLTHG    & $0.002$  & $0.059$  & $0.100$  & $0.011$  & $-$  \\
HLTHF    & $0.294$  & $0.209$  & $0.358$  & $0.411$  & $-$  \\
HLTHP    & $0.801$  & $0.588$  & $0.613$  & $0.700$  & $-$  \\
MHI      & $-0.039$ & $-0.043$ & $-0.029$ & $-0.060$ & $-0.053$ \\
$\textbf{b}_{11}$   & $\mathbf{3.107} \; (0.015)$  & $\mathbf{3.590} \; (0.020)$  & $\mathbf{4.353} \; (0.025)$  & $\mathbf{5.557} \; (0.020)$  & $\mathbf{6.771} \; (0.008)$  \\
$\textbf{b}_{12}$   & $\mathbf{1.700} \; (0.016)$  & $\mathbf{2.362} \; (0.020)$  & $\mathbf{3.285} \; (0.026)$  & $\mathbf{4.044} \; (0.018)$  & $\mathbf{4.705} \; (0.007)$  \\
$\sigma_\tau$    & $\mathbf{0.187} \; (0.001)$  & $\mathbf{0.355} \; (0.003)$  & $\mathbf{0.471} \; (0.004)$  & $\mathbf{0.372} \; (0.003)$  & $\mathbf{0.187} \; (0.001)$  \\
$\lambda_\tau$ & $0.013$  & $0.018$  & $0.018$  & $0.032$  & $0.566$  \\
\hline

Panel C: \\
$\pi_1$ & $0.495 \; (0.006)$ & $0.484 \; (0.008)$ & $0.487 \; (0.009)$ & $\mathbf{0.337} \; (0.008)$ & $\mathbf{0.328} \; (0.007)$ \\
$\pi_2$ & $0.505 \; (0.006)$ & $0.516 \; (0.008)$ & $0.513 \; (0.009)$ & $\mathbf{0.663} \; (0.008)$ & $\mathbf{0.672} \; (0.007)$ \\
\hline

Panel D: \\
$\log(L)$ & -10349.09 & -7446.91 & -6742.39 & -9394.92 & -12392.29 \\ 
$\#$ par & 36 & 35 & 36 & 36 & 29 \\ 
AIC & 20770.18 & 14963.82 & 13556.77 & 18861.84 & 24844.59 \\ 
BIC & 21055.04 & 15240.77 & 13841.63 & 19146.70 & 25081.97 \\ 
\hline

\end{tabular}}
\caption{\footnotesize Penalized two-part finite mixture of quantile regressions coefficient estimates. Panel A refers to the binary part while Panel B refers to the positive part of the model for the investigated quantile levels. 
Panel C illustrates the estimated mixing probabilities. Panel D reports the log-likelihood, number of nonzero model parameters and penalized likelihood criteria (AIC, BIC).}
\label{tab:pen}
\end{table}

\begin{table}[h]
\centering
 \smallskip 
 \resizebox{1.0\columnwidth}{!}{
 \setlength\tabcolsep{13pt}
\begin{tabular}{l c c c c c }
\hline
Covariate &\multicolumn{5}{c}{$\tau$-th quantile} \\\cmidrule(r){2-6}
  & 10 & 25 & 50 & 75 & 90 \\ 
\hline

LOGC     & $\mathbf{-0.202} \; (0.026)$ & $\mathbf{-0.159} \; (0.017)$ & $\mathbf{-0.147} \; (0.017)$ & $\mathbf{-0.100} \; (0.025)$ & $\mathbf{-0.102} \; (0.026)$ \\
LFAM     & $\mathbf{-0.130} \; (0.023)$ & $\mathbf{-0.108} \; (0.018)$ & $\mathbf{-0.116} \; (0.019)$ & $-0.039 \; (0.030)$          & $\mathbf{-0.061} \; (0.025)$ \\
LINC     & $\mathbf{0.062} \; (0.028)$  & $\mathbf{0.066} \; (0.022)$  & $\mathbf{0.055} \; (0.018)$  & $\mathbf{0.079} \; (0.035)$  & $0.055 \; (0.034)$           \\
XAGE     & $\mathbf{0.232} \; (0.028)$  & $\mathbf{0.169} \; (0.023)$  & $\mathbf{0.173} \; (0.024)$  & $\mathbf{0.115} \; (0.029)$  & $\mathbf{0.110} \; (0.035)$  \\
FEMALE   & $\mathbf{0.328} \; (0.036)$  & $\mathbf{0.244} \; (0.029)$  & $\mathbf{0.280} \; (0.031)$  & $\mathbf{0.377} \; (0.041)$  & $\mathbf{0.368} \; (0.039)$  \\
CHILD    & $\mathbf{0.064} \; (0.033)$  & $-0.020 \; (0.030)$          & $\mathbf{-0.083} \; (0.036)$ & $\mathbf{-0.279} \; (0.040)$ & $\mathbf{-0.242} \; (0.042)$ \\
FEMCHILD & $\mathbf{-0.254} \; (0.035)$ & $\mathbf{-0.281} \; (0.033)$ & $\mathbf{-0.266} \; (0.037)$ & $\mathbf{-0.464} \; (0.042)$ & $\mathbf{-0.478} \; (0.044)$ \\
BLACK    & $\mathbf{-0.192} \; (0.035)$ & $\mathbf{-0.359} \; (0.039)$ & $\mathbf{-0.315} \; (0.042)$ & $\mathbf{-0.290} \; (0.050)$ & $\mathbf{-0.305} \; (0.055)$ \\
EDUCDEC  & $0.056 \; (0.046)$           & $\mathbf{0.051} \; (0.016)$  & $0.023 \; (0.017)$           & $-0.016 \; (0.028)$          & $-0.011 \; (0.027)$          \\
PHYSLM   & $\mathbf{0.326} \; (0.037)$  & $\mathbf{0.274} \; (0.042)$  & $\mathbf{0.342} \; (0.046)$  & $\mathbf{0.428} \; (0.049)$  & $\mathbf{0.450} \; (0.045)$  \\
DISEA    & $\mathbf{0.103} \; (0.030)$  & $\mathbf{0.150} \; (0.022)$  & $\mathbf{0.115} \; (0.020)$  & $\mathbf{0.057} \; (0.023)$  & $\mathbf{0.074} \; (0.024)$  \\
HLTHG    & $0.024 \; (0.044)$           & $\mathbf{0.095} \; (0.030)$  & $\mathbf{0.073} \; (0.033)$  & $\mathbf{0.087} \; (0.042)$  & $\mathbf{0.148} \; (0.044)$  \\
HLTHF    & $\mathbf{0.268} \; (0.038)$  & $\mathbf{0.357} \; (0.042)$  & $\mathbf{0.327} \; (0.049)$  & $\mathbf{0.332} \; (0.043)$  & $\mathbf{0.272} \; (0.047)$  \\
HLTHP    & $\mathbf{0.573} \; (0.026)$  & $\mathbf{0.631} \; (0.046)$  & $\mathbf{0.657} \; (0.040)$  & $\mathbf{0.866} \; (0.038)$  & $\mathbf{0.698} \; (0.037)$  \\
MHI      & $-0.013 \; (0.024)$          & $\mathbf{-0.072} \; (0.018)$ & $\mathbf{-0.037} \; (0.015)$ & $\mathbf{-0.080} \; (0.025)$ & $\mathbf{-0.077} \; (0.023)$ \\
Intercept   & $\mathbf{2.497} \; (0.036)$  & $\mathbf{3.153} \; (0.024)$  & $\mathbf{3.845} \; (0.027)$  & $\mathbf{4.637} \; (0.034)$  & $\mathbf{4.833} \; (0.030)$  \\
$\sigma_\tau$    & $\mathbf{0.155} \; (0.001)$  & $\mathbf{0.321} \; (0.003)$  & $\mathbf{0.430} \; (0.004)$  & $\mathbf{0.340} \; (0.003)$  & $\mathbf{0.291} \; (0.003)$  \\
\hline

\end{tabular}}
\caption{
\footnotesize LQMM coefficient estimates for the positive outcomes for the investigated quantile levels. 
Standard errors are in parentheses. Parameter estimates are displayed in boldface when significant at the standard 5\% level.}
\label{tab:unpen}
\end{table}

\section{Conclusions}\label{sec:con}
This paper introduces a two-part finite mixture of quantile regressions for mixed-type outcomes under a longitudinal setting. Random effect coefficients are added in both the binary and the positive decision mechanisms to account for zero inflation and unobserved heterogeneity. Rather than assuming a parametric distribution on the random coefficients distribution, we approximate it using a multivariate discrete variable defined on a finite number of support points. 
Estimation of the model parameters is based on a suitable likelihood-based EM algorithm. In addition, a LASSO penalized version of the algorithm is proposed as an automatic data-driven procedure to perform variable selection. The application of the proposed method to the RHIE on health behaviors and attitudes shows consistent results with existing studies.\\ 
The proposed approach can be further extended to consider time-varying sources of unobserved heterogeneity via individual-specific coefficients evolving according a hidden Markov chain. Secondly, we may extend the univariate quantile framework to a multivariate quantile regression setting taking into account for the correlation among the marginals of a multivariate response variable (\cite{petrella2019joint}).



\bibliography{biblio}

\begin{thebibliography}{57}
\providecommand{\natexlab}[1]{#1}
\providecommand{\url}[1]{\texttt{#1}}
\expandafter\ifx\csname urlstyle\endcsname\relax
  \providecommand{\doi}[1]{doi: #1}\else
  \providecommand{\doi}{doi: \begingroup \urlstyle{rm}\Url}\fi

\bibitem[Alf\`{o} and Maruotti(2010)]{Alfo2010}
{\rm Alf\`{o}, M. {\rm and} Maruotti, A.} (2010).
\newblock Two-part regression models for longitudinal zero-inflated count data.
\newblock \emph{Canadian Journal of Statistics}, {\bf 38}\penalty0 (2),
  \penalty0 197--216.

\bibitem[Alf{\`o} et~al.(2017)Alf{\`o}, Salvati, and Ranallli]{alfo2017finite}
{\rm Alf{\`o}, M., Salvati, N., {\rm and} Ranallli, M.~G.} (2017).
\newblock Finite mixtures of quantile and {M}-quantile regression models.
\newblock \emph{Statistics and Computing}, {\bf 27}\penalty0 (2), \penalty0
  547--570.

\bibitem[Alhamzawi et~al.(2012)Alhamzawi, Yu, and
  Benoit]{alhamzawi2012bayesian}
{\rm Alhamzawi, R., Yu, K., {\rm and} Benoit, D.~F.} (2012).
\newblock Bayesian adaptive {L}asso quantile regression.
\newblock \emph{Statistical Modelling}, {\bf 12}\penalty0 (3), \penalty0
  279--297.

\bibitem[Bago~d'Uva(2006)]{bago2006latent}
{\rm Bago~d'Uva, T.} (2006).
\newblock Latent class models for utilisation of health care.
\newblock \emph{Health Economics}, {\bf 15}\penalty0 (4), \penalty0 329--343.

\bibitem[Bassett and Chen(2002)]{Bassett2002}
{\rm Bassett, G.~W. {\rm and} Chen, H.-L.} (2002).
\newblock \emph{Portfolio style: Return-based attribution using quantile
  regression}, pages 293--305.
\newblock Physica-Verlag HD, Heidelberg.
\newblock ISBN 978-3-662-11592-3.
\newblock \doi{10.1007/978-3-662-11592-3_15}.

\bibitem[Basu and Manning(2009)]{Basu2009}
{\rm Basu, A. {\rm and} Manning, W.~G.} (2009).
\newblock Issues for the next generation of health care cost analyses.
\newblock \emph{Medical Care}, {\bf 47}, \penalty0 S109--S114.

\bibitem[Belasco and Ghosh(2012)]{belasco2012modelling}
{\rm Belasco, E.~J. {\rm and} Ghosh, S.~K.} (2012).
\newblock Modelling semi-continuous data using mixture regression models with
  an application to cattle production yields.
\newblock \emph{The Journal of Agricultural Science}, {\bf 150}\penalty0 (1),
  \penalty0 109--121.

\bibitem[Belotti et~al.(2015)Belotti, Deb, Manning, and
  Norton]{belotti2015twopm}
{\rm Belotti, F., Deb, P., Manning, W.~G., {\rm and} Norton, E.~C.} (2015).
\newblock Twopm: Two-part models.
\newblock \emph{The Stata Journal}, {\bf 15}\penalty0 (1), \penalty0 3--20.

\bibitem[Bernardi et~al.(2015)Bernardi, Gayraud, Petrella,
  et~al.]{bernardi2015bayesian}
{\rm Bernardi, M., Gayraud, G., Petrella, L., et~al.} (2015).
\newblock Bayesian tail risk interdependence using quantile regression.
\newblock \emph{Bayesian Analysis}, {\bf 10}\penalty0 (3), \penalty0 553--603.

\bibitem[Bernardi et~al.(2018)Bernardi, Bottone, and
  Petrella]{bernardi2018bayesian}
{\rm Bernardi, M., Bottone, M., {\rm and} Petrella, L.} (2018).
\newblock Bayesian quantile regression using the skew exponential power
  distribution.
\newblock \emph{Computational Statistics \& Data Analysis}, {\bf 126},
  \penalty0 92--111.

\bibitem[Biswas et~al.(2020)Biswas, Ghosh, and Das]{biswas2020semi}
{\rm Biswas, J., Ghosh, P., {\rm and} Das, K.} (2020).
\newblock A semi-parametric quantile regression approach to zero-inflated and
  incomplete longitudinal outcomes.
\newblock \emph{AStA Advances in Statistical Analysis}, pages 1--23.

\bibitem[Cole and Green(1992)]{cole1992smoothing}
{\rm Cole, T.~J. {\rm and} Green, P.~J.} (1992).
\newblock Smoothing reference centile curves: the {LMS} method and penalized
  likelihood.
\newblock \emph{Statistics in Medicine}, {\bf 11}\penalty0 (10), \penalty0
  1305--1319.

\bibitem[Deb and Trivedi(2002)]{deb2002structure}
{\rm Deb, P. {\rm and} Trivedi, P.~K.} (2002).
\newblock The structure of demand for health care: latent class versus two-part
  models.
\newblock \emph{Journal of Health Economics}, {\bf 21}\penalty0 (4), \penalty0
  601--625.

\bibitem[Diehr et~al.(1999)Diehr, Yanez, Ash, Hornbrook, and
  Lin]{diehr1999methods}
{\rm Diehr, P., Yanez, D., Ash, A., Hornbrook, M., {\rm and} Lin, D.} (1999).
\newblock Methods for analyzing health care utilization and costs.
\newblock \emph{Annual Review of Public Health}, {\bf 20}\penalty0 (1),
  \penalty0 125--144.

\bibitem[Duan et~al.(1983)Duan, Manning, Morris, and Newhouse]{Duan1983}
{\rm Duan, N., Manning, W.~G., Morris, C.~N., {\rm and} Newhouse, J.~P.}
  (1983).
\newblock A comparison of alternative models for the demand for medical care.
\newblock \emph{Journal of Business \& Economic Statistics}, {\bf 1}\penalty0
  (2), \penalty0 115--126.

\bibitem[Fan and Lv(2010)]{fan2010selective}
{\rm Fan, J. {\rm and} Lv, J.} (2010).
\newblock A selective overview of variable selection in high dimensional
  feature space.
\newblock \emph{Statistica Sinica}, {\bf 20}\penalty0 (1), \penalty0 101.

\bibitem[Farcomeni and Viviani(2015)]{farcomeni2015longitudinal}
{\rm Farcomeni, A. {\rm and} Viviani, S.} (2015).
\newblock Longitudinal quantile regression in the presence of informative
  dropout through longitudinal-survival joint modeling.
\newblock \emph{Statistics in Medicine}, {\bf 34}\penalty0 (7), \penalty0
  1199--1213.

\bibitem[Farewell et~al.(2017)Farewell, Long, Tom, Yiu, and
  Su]{farewell2017two}
{\rm Farewell, V., Long, D., Tom, B., Yiu, S., {\rm and} Su, L.} (2017).
\newblock Two-part and related regression models for longitudinal data.
\newblock \emph{Annual Review of Statistics and Its Application}, {\bf 4},
  \penalty0 283--315.

\bibitem[Geraci and Bottai(2014)]{Geraci2014}
{\rm Geraci, M. {\rm and} Bottai, M.} (2014).
\newblock Linear quantile mixed models.
\newblock \emph{Statistics and Computing}, {\bf 24}\penalty0 (3), \penalty0
  461--479.

\bibitem[Green(1990)]{green1990use}
{\rm Green, P.~J.} (1990).
\newblock On use of the {EM} algorithm for penalized likelihood estimation.
\newblock \emph{Journal of the Royal Statistical Society: Series B
  (Methodological)}, {\bf 52}\penalty0 (3), \penalty0 443--452.

\bibitem[Grilli et~al.(2016)Grilli, Rampichini, and Varriale]{Grilli2016}
{\rm Grilli, L., Rampichini, C., {\rm and} Varriale, R.} (2016).
\newblock Statistical modelling of gained university credits to evaluate the
  role of pre-enrolment assessment tests: An approach based on quantile
  regression for counts.
\newblock \emph{Statistical Modelling}, {\bf 16}\penalty0 (1), \penalty0
  47--66.

\bibitem[Hendricks and Koenker(1992)]{hendricks1992hierarchical}
{\rm Hendricks, W. {\rm and} Koenker, R.} (1992).
\newblock Hierarchical spline models for conditional quantiles and the demand
  for electricity.
\newblock \emph{Journal of the American Statistical Association}, {\bf
  87}\penalty0 (417), \penalty0 58--68.

\bibitem[Heras et~al.(2018)Heras, Moreno, and
  Vilar-Zan{\'o}n]{heras2018application}
{\rm Heras, A., Moreno, I., {\rm and} Vilar-Zan{\'o}n, J.~L.} (2018).
\newblock An application of two-stage quantile regression to insurance
  ratemaking.
\newblock \emph{Scandinavian Actuarial Journal}, {\bf 2018}\penalty0 (9),
  \penalty0 753--769.

\bibitem[Iversen et~al.(2015)Iversen, Aas, Rosenqvist, Hakkinen, and on~behalf
  of the EuroHOPE~study group]{Iversen2015}
{\rm Iversen, T., Aas, E., Rosenqvist, G., Hakkinen, U., {\rm and} on~behalf of
  the EuroHOPE~study group} (2015).
\newblock Comparative analysis of treatment costs in {EUROHOPE}.
\newblock \emph{Health Economics}, {\bf 24}, \penalty0 5--22.

\bibitem[Koenker(2005)]{koenker2005}
{\rm Koenker, R.} (2005).
\newblock \emph{Quantile Regression}.
\newblock Cambridge University Press.

\bibitem[Koenker and Bassett(1978)]{koenker1978}
{\rm Koenker, R. {\rm and} Bassett, G.} (1978).
\newblock Regression quantiles.
\newblock \emph{Econometrica: Journal of the Econometric Society}, {\bf
  46}\penalty0 (1), \penalty0 33--50.

\bibitem[Koenker et~al.(2017)Koenker, Chernozhukov, He, and
  Peng]{koenker2017handbook}
{\rm Koenker, R., Chernozhukov, V., He, X., {\rm and} Peng, L.} (2017).
\newblock \emph{Handbook of quantile regression}.
\newblock CRC press.

\bibitem[Kozumi and Kobayashi(2011)]{kozumi2011gibbs}
{\rm Kozumi, H. {\rm and} Kobayashi, G.} (2011).
\newblock Gibbs sampling methods for bayesian quantile regression.
\newblock \emph{Journal of Statistical Computation and Simulation}, {\bf
  81}\penalty0 (11), \penalty0 1565--1578.

\bibitem[Laird(1978)]{laird1978nonparametric}
{\rm Laird, N.} (1978).
\newblock Nonparametric maximum likelihood estimation of a mixing distribution.
\newblock \emph{Journal of the American Statistical Association}, {\bf
  73}\penalty0 (364), \penalty0 805--811.

\bibitem[Laporta et~al.(2018)Laporta, Merlo, and
  Petrella]{laporta2018selection}
{\rm Laporta, A.~G., Merlo, L., {\rm and} Petrella, L.} (2018).
\newblock Selection of {V}alue at {R}isk models for energy commodities.
\newblock \emph{Energy Economics}, {\bf 74}, \penalty0 628--643.

\bibitem[Liu(2009)]{liu2009joint}
{\rm Liu, L.} (2009).
\newblock Joint modeling longitudinal semi-continuous data and survival, with
  application to longitudinal medical cost data.
\newblock \emph{Statistics in Medicine}, {\bf 28}\penalty0 (6), \penalty0
  972--986.

\bibitem[Manning et~al.(1987)Manning, Newhouse, Duan, Keeler, and
  Leibowitz]{manning1987health}
{\rm Manning, W.~G., Newhouse, J.~P., Duan, N., Keeler, E.~B., {\rm and}
  Leibowitz, A.} (1987).
\newblock Health insurance and the demand for medical care: evidence from a
  randomized experiment.
\newblock \emph{The American Economic Review}, pages 251--277.

\bibitem[Marino and Farcomeni(2015)]{marino2015linear}
{\rm Marino, M.~F. {\rm and} Farcomeni, A.} (2015).
\newblock Linear quantile regression models for longitudinal experiments: an
  overview.
\newblock \emph{Metron}, {\bf 73}\penalty0 (2), \penalty0 229--247.

\bibitem[Marino et~al.(2018)Marino, Tzavidis, and Alf{\`o}]{marino2018mixed}
{\rm Marino, M.~F., Tzavidis, N., {\rm and} Alf{\`o}, M.} (2018).
\newblock Mixed hidden {M}arkov quantile regression models for longitudinal
  data with possibly incomplete sequences.
\newblock \emph{Statistical Methods in Medical Research}, {\bf 27}\penalty0
  (7), \penalty0 2231--2246.

\bibitem[Maruotti and Raponi(2014)]{Maruotti2014}
{\rm Maruotti, A. {\rm and} Raponi, V.} (2014).
\newblock On baseline conditions for zero-inflated longitudinal count data.
\newblock \emph{Communications in Statistics - Simulation and Computation},
  {\bf 43}\penalty0 (4), \penalty0 743--760.

\bibitem[Mihaylova et~al.(2011)Mihaylova, Briggs, O'Hagan, and
  Thompson]{mihaylova2011review}
{\rm Mihaylova, B., Briggs, A., O'Hagan, A., {\rm and} Thompson, S.~G.} (2011).
\newblock Review of statistical methods for analysing healthcare resources and
  costs.
\newblock \emph{Health Economics}, {\bf 20}\penalty0 (8), \penalty0 897--916.

\bibitem[Min and Agresti(2002)]{min2002modeling}
{\rm Min, Y. {\rm and} Agresti, A.} (2002).
\newblock Modeling nonnegative data with clumping at zero: a survey.
\newblock \emph{Journal of the Iranian Statistical Society}, {\bf 1}\penalty0
  (1), \penalty0 7--33.

\bibitem[Min and Agresti(2005)]{min2005random}
{\rm Min, Y. {\rm and} Agresti, A.} (2005).
\newblock Random effect models for repeated measures of zero-inflated count
  data.
\newblock \emph{Statistical Modelling}, {\bf 5}\penalty0 (1), \penalty0 1--19.

\bibitem[Mullahy(1986)]{Mullahy1986}
{\rm Mullahy, J.} (1986).
\newblock Specification and testing of some modified count data models.
\newblock \emph{Journal of Econometrics}, {\bf 33}\penalty0 (3), \penalty0 341
  -- 365.

\bibitem[Neelon et~al.(2016{\natexlab{a}})Neelon, O'Malley, and
  Smith]{neelon2016modeling1}
{\rm Neelon, B., O'Malley, A.~J., {\rm and} Smith, V.~A.} (2016{\natexlab{a}}).
\newblock Modeling zero-modified count and semicontinuous data in health
  services research part 1: background and overview.
\newblock \emph{Statistics in Medicine}, {\bf 35}\penalty0 (27), \penalty0
  5070--5093.

\bibitem[Neelon et~al.(2016{\natexlab{b}})Neelon, O'Malley, and
  Smith]{neelon2016modeling2}
{\rm Neelon, B., O'Malley, A.~J., {\rm and} Smith, V.~A.} (2016{\natexlab{b}}).
\newblock Modeling zero-modified count and semicontinuous data in health
  services research part 2: case studies.
\newblock \emph{Statistics in Medicine}, {\bf 35}\penalty0 (27), \penalty0
  5094--5112.

\bibitem[Pandey and Nguyen(1999)]{pandey1999comparative}
{\rm Pandey, G.~R. {\rm and} Nguyen, V.-T.-V.} (1999).
\newblock A comparative study of regression based methods in regional flood
  frequency analysis.
\newblock \emph{Journal of Hydrology}, {\bf 225}\penalty0 (1-2), \penalty0
  92--101.

\bibitem[Petrella and Raponi(2019)]{petrella2019joint}
{\rm Petrella, L. {\rm and} Raponi, V.} (2019).
\newblock Joint estimation of conditional quantiles in multivariate linear
  regression models with an application to financial distress.
\newblock \emph{Journal of Multivariate Analysis}, {\bf 173}, \penalty0 70--84.

\bibitem[Petrella et~al.(2018)Petrella, Laporta, and Merlo]{petrella2018cross}
{\rm Petrella, L., Laporta, A.~G., {\rm and} Merlo, L.} (2018).
\newblock Cross-{C}ountry {A}ssessment of {S}ystemic {R}isk in the {E}uropean
  {S}tock {M}arket: {E}vidence from a {C}o{V}a{R} {A}nalysis.
\newblock \emph{Social Indicators Research}, pages 1--18.
\newblock \doi{10.1007/s11205-018-1881-8}.

\bibitem[Rabe-Hesketh et~al.(2005)Rabe-Hesketh, Skrondal, and
  Pickles]{rabe2005maximum}
{\rm Rabe-Hesketh, S., Skrondal, A., {\rm and} Pickles, A.} (2005).
\newblock Maximum likelihood estimation of limited and discrete dependent
  variable models with nested random effects.
\newblock \emph{Journal of Econometrics}, {\bf 128}\penalty0 (2), \penalty0
  301--323.

\bibitem[Reich et~al.(2011)Reich, Fuentes, and Dunson]{reich2011bayesian}
{\rm Reich, B.~J., Fuentes, M., {\rm and} Dunson, D.~B.} (2011).
\newblock Bayesian spatial quantile regression.
\newblock \emph{Journal of the American Statistical Association}, {\bf
  106}\penalty0 (493), \penalty0 6--20.

\bibitem[Royston and Altman(1994)]{royston1994regression}
{\rm Royston, P. {\rm and} Altman, D.~G.} (1994).
\newblock Regression using fractional polynomials of continuous covariates:
  parsimonious parametric modelling.
\newblock \emph{Journal of the Royal Statistical Society: Series C (Applied
  Statistics)}, {\bf 43}\penalty0 (3), \penalty0 429--453.

\bibitem[Sauzet et~al.(2019)Sauzet, Razum, Widera, and Brzoska]{sauzet2019two}
{\rm Sauzet, O., Razum, O., Widera, T., {\rm and} Brzoska, P.} (2019).
\newblock Two-part models and quantile regression for the analysis of survey
  data with a spike. {T}he example of satisfaction with health care.
\newblock \emph{Frontiers in Public Health}, {\bf 7}, \penalty0 146.

\bibitem[Schwarz et~al.(1978)]{schwarz1978estimating}
{\rm Schwarz, G. et~al.} (1978).
\newblock Estimating the dimension of a model.
\newblock \emph{The Annals of Statistics}, {\bf 6}\penalty0 (2), \penalty0
  461--464.

\bibitem[Tian et~al.(2018)Tian, Gao, and Yang]{tian2018quantile}
{\rm Tian, F., Gao, J., {\rm and} Yang, K.} (2018).
\newblock A quantile regression approach to panel data analysis of health-care
  expenditure in organisation for economic co-operation and development
  countries.
\newblock \emph{Health Economics}, {\bf 27}\penalty0 (12), \penalty0
  1921--1944.

\bibitem[Tian et~al.(2014)Tian, Tian, and Zhu]{tian2014linear}
{\rm Tian, Y., Tian, M., {\rm and} Zhu, Q.} (2014).
\newblock Linear quantile regression based on {EM} algorithm.
\newblock \emph{Communications in Statistics-Theory and Methods}, {\bf
  43}\penalty0 (16), \penalty0 3464--3484.

\bibitem[Tibshirani(1996)]{Tibshirani1996}
{\rm Tibshirani, R.} (1996).
\newblock Regression shrinkage and selection via the {L}asso.
\newblock \emph{Journal of the Royal Statistal Society. Series B}, {\bf 58},
  \penalty0 267--288.

\bibitem[Waldmann(2018)]{waldmann2018quantile}
{\rm Waldmann, E.} (2018).
\newblock Quantile regression: a short story on how and why.
\newblock \emph{Statistical Modelling}, {\bf 18}\penalty0 (3-4), \penalty0
  203--218.

\bibitem[Wasserman and Roeder(2009)]{wasserman2009high}
{\rm Wasserman, L. {\rm and} Roeder, K.} (2009).
\newblock High dimensional variable selection.
\newblock \emph{Annals of Statistics}, {\bf 37}\penalty0 (5A), \penalty0 2178.

\bibitem[Winkelmann(2004)]{winkelmann2004health}
{\rm Winkelmann, R.} (2004).
\newblock Health care reform and the number of doctor visits—an econometric
  analysis.
\newblock \emph{Journal of Applied Econometrics}, {\bf 19}\penalty0 (4),
  \penalty0 455--472.

\bibitem[Yu and Moyeed(2001)]{yu2001bayesian}
{\rm Yu, K. {\rm and} Moyeed, R.~A.} (2001).
\newblock Bayesian quantile regression.
\newblock \emph{Statistics \& Probability Letters}, {\bf 54}\penalty0 (4),
  \penalty0 437--447.

\bibitem[Zhou(2002)]{Zhou2002}
{\rm Zhou, X.-H.} (2002).
\newblock Inferences about population means of health care costs.
\newblock \emph{Statistical Methods in Medical Research}, {\bf 11}\penalty0
  (4), \penalty0 327--339.

\end{thebibliography}

\end{document}